# Detecting Antibody-Antigen Interactions with Chiral Plasmons: Factors Influencing Chiral Plasmonic Sensing.

D. Koyroytsaltis-McQuire[1], C. Gilroy[1], L. Barron[1], N. Gadegaard[2], A. Karimullah[1] and M. Kadodwala[1*]

[1] School of Chemistry, University of Glasgow, Glasgow, G12 8QQ, UK
[2] School of Engineering, Rankine Building, University of Glasgow, Glasgow G12 8LT, U.K

**Abstract**

Chiral near fields possessing enhanced asymmetry (superchirality), created by the interaction of light with (chiral) nanostructures, potentially provide a route to novel sensing and metrology technologies for biophysical applications. However, the mechanisms by which these near fields lead to the detection of chiral media is still poorly understood. Using a combination of numerical modelling and experimental measurements on an antibody-antigen exemplar system we illustrate important factors that influence the efficacy of chiral sensing. We demonstrate that localised and lattice chiral resonances display enantiomeric sensitivity. However, only the localised resonances exhibit strong dependency on the structure of the chiral media detected. This can be attributed to the ability of birefringent chiral layers to strongly modify the properties of near fields by acting as a sink / source of optical chirality, and hence alter inductive coupling between nanostructure elements. In addition, we highlight how surface morphology / defects may amplify sensing capabilities of localised chiral plasmonic modes by mediating inductive coupling.

**Introduction**

Spectroscopic techniques based on the differential interaction of circularly polarised light, such as circular dichroism, can provide a rapid method for the detection and low-resolution structural characterisation of biologically relevant molecular materials[1]. The inherent weakness of the optically active response intrinsically limits the sensitivity of chiroptical spectroscopic methods, with maximum detection sensitivities typically at the $\geq \mu g$ level. It has been proposed that the sensitivities of chiroptical spectroscopies can be amplified using electromagnetic (EM) fields which in highly localised regions of space can have greater chiral asymmetries than circularly polarised light (CPL), a property sometimes referred to as superchirality[2, 3]. The chiral asymmetry of EM fields is parameterised using optical chirality density $(C)$[4] typically normalised against the value for the equivalent CPL. Near fields created by light scattering from nanostructures can have $|C| > 1$[5], this has been demonstrated using chiral plasmonic[5-8] and achiral dielectric nanostructures[9-11]. Introducing chiral media into the near field regions of chiral nanostructure can lead to asymmetric changes in the chiroptical response measured in the far field. This phenomenon offers an appealing route to novel ultrasensitive biosensing technologies with ≤ pg detection limits[8, 12-19]. For this phenomenon to be exploited effectively requires an understanding of chiral light - matter interactions. The crucial issues to be addressed are: how the introduction of chiral media into the near field region of nanostructures leads to a significant asymmetry in a far-field chiroptical response; and is the detection phenomena generic or are there constraints placed on the nature of the types of chiral media that can be detected? Through the combinationof both numerical modelling and experimental measurement. we have determined whether chiral resonances of the gammadion nanostructures are localised or lattice modes. Numerical simulations indicate that chiral sensing capabilities of lattice modes do not show a strong dependence on the structure of the chiral dielectric. In contrast localised modes are far more sensitive to the structure of the chiral dielectric. Specifically, birefringence induced by anisotropic ordering of chiral layers (20 nm thick) significantly amplifies dissymmetry between enantiomorphic structures in the chiroptical response of localised resonances.

The ability of chiral layers to induce asymmetric changes in the chiroptical responses of enantiomorphic structures is associated with an ability to induce differential changes in the properties of near fields. In the absence of chiral media, symmetry equivalent combinations of light circular polarisation and nanostructure handedness, near fields have opposite signs of optical chirality but are otherwise identical. Introduction of chiral layers breaks this relationship, with the largest divergence occurring when the chiral layers possess birefringence. This effect is attributed to the ability of chiral birefringent layers to act as a sink (source) of optical chirality. Consequently, compared to isotropic chiral media, birefringent layers induce greater asymmetries in near field properties, resulting in a greater change in far field optical response.

**Background**

Optical chirality (C),

$$C \equiv \frac{\varepsilon_0}{2} \boldsymbol{E} \cdot \nabla \times \boldsymbol{E} + \frac{1}{2\mu_0} \boldsymbol{B} \cdot \nabla \times \boldsymbol{B} \qquad (1)$$

is a conserved property of light[4, 20, 21], like energy, and is equivalent to optical spin-density. When chiral EM fields interact with chiral matter, optical chirality can be either exchanged or dissipated through absorption. Thus, the optical chirality flux of a light beam can be changed in a chiral light - matter interaction[22-25]. These processes depend on the handedness of both the circular polarisation of light and the media. The differential absorption of CPL (dissipation of optical chirality) by chiral media is the basis of the chiroptical technique circular dichroism. Optical chirality can be exchanged between CPL and a medium without absorption. For instance, optical chirality can be transferred to a non-absorbing birefringent material resulting in the depolarisation of the CPL beam. Alternatively, linearly polarised light can also become elliptically polarised (*i.e.* gains optical chirality) by passing through birefringent materials. The transfer of optical spin angular momentum from CPL sufficient to create an opto-mechanical torque to rotate macroscopic objects was first demonstrate by Beth in 1936[26].

The central premise of this study is that chiral birefringent layers act as efficient sinks of near field optical chirality. This causes significant divergence in the reciprocity of the *C* and intensities of fields possessed by left (LH) and right-handed (RH) nanostructures. Consequently, this causes asymmetric changes in the chiroptical properties of LH and RH structures, measured in the far field, which enhances chiral sensing capabilities.

**Results**

A gammadion has four-fold rotational symmetry and in free space belongs to the $C_{4h}$ point group. When placed on to a surface, mirror symmetry is broken, and it becomes chiral with a point group symmetry of $C_4$. Metamaterials consisting of periodic arrays of gammadia display large levels of optical activity in the visible and near IR region of the spectrum[27-30]. Gammadia based metamaterials of both pure enantiomorphs and racemic structures have been used for the detection of chiral molecular materials[8, 28]. Experimental circular dichroism (CD) spectra from LH and RH gammadion structures immersed in buffer solution are compared with those derived from numerical simulations, in **figure 1**. The latter are based on an idealised model of the gammadia in water, using the finite element method, and replicate the experimental data reasonably well. The magnitude of the simulated CD spectra is ~ 1 order greater than that observed experimentally. This in part can be accounted for by the checkerboard structure of the substrate used in the experiment. To reduce fabrication time, only 50% of the area is covered in nanostructures. In addition, the highly idealised nature of the gammadion structure used in the simulation, **figure 2**, which does not account for surface roughness or defects (*e.g.* missing or damaged structures) can in part be used to rationalise further differences with experiment. The modelling results can be compared to numerical simulations on a similar structure performed by Phua and co-workers[31], in which the finite difference time-domain method was used. As in previous studies[28, 31, 32] we focus on three resonances, which have been labelled I, II and III. Field maps for the three resonances showing both electric field intensities and optical chirality are shown in **figure 3.** The fields associated with modes I and II are localised on the nanostructure, with differing field distributions and intensities. The origin of these two modes can be interpreted with the Born-Kuhn (coupled oscillator) model of optical activity[33-35]. Within this framework modes I and II can be considered to originate from out- and in-phase combinations of the coupled oscillators system. Applying the Born-Kuhn model to the gammadion structure, the two orthogonal rods which make up each of the four arms, can be approximated to two harmonic oscillators, which can conductively and inductively couple.

Mode III is located a wavelength, 780 nm. close to the periodicity of the structure; and is associated with a field distribution which has intense regions between gammadia. Both factors point to mode III being a Bloch (surface lattice) mode[36].

A group theory analysis based on the ($C_4$) point group symmetry of the gammadion provides information on the non-Bloch modes localised on the structure. The symmetry analysis is based on considering the gammadion structure to consist of 8 rods each being assigned a vector (representing a dipole moment). Using this as a basis it can be determined that there are 6 modes, 2A+2B+2E, with only the doubly degenerate E modes contributing to CD spectra. The two E modes can then be assigned to the two out- / in-phase combinations, modes I and II respectively. Thus, the symmetry analysis is consistent with the predictions of the coupled oscillator model and the numerical simulations.

The mode assignments made above are also supported by experimental observation. We have collected spectra from the two enantiomorphic structures which have been given a gradually increasing incline, shown in **figure 4**. As expected, the Bloch / lattice mode (mode III) splits in to two components, which arise because the x and y directions (orthogonal to the z-direction of propagation) are no longer equivalent, as the light beam has a different angle of incidence for each. However, the localised modes (I and II) are not significantly affected. Bisignate line shapes in CD are signatures of the couple oscillator model of chirality. Individually mode I and II would each give a bisignate line shape, albeit with opposite phases. Consequently, when sufficiently close in wavelength the bisignate forms of the two resonances overlap giving the characteristic "W" line shape observed in the 550- 700 nm range. The relative contribution of each bisignate changes, with an increase contribution of mode II (in-phase combination) with increasing incidence angle. This results in a red shift in the maximum of the peak between resonances I and II with increasing inclination angle.

**Parameterising spectral asymmetry.**

The ability to detect chiral (bio)materials with chiral metamaterials is based on the premise that they asymmetrically change the optical properties of enantiomorphic structures. These asymmetries can manifest as differential shifts in the positions of resonances which can be parameterised by:

$$\Delta\Delta\lambda = \Delta\lambda_R - \Delta\lambda_L \qquad (2)$$

where $\Delta\lambda_{L/R}$ are shifts in the position of resonances (I, II, II) in the presence of the chiral media, relative to an achiral reference, which in this case is buffer solution. In addition the presence of chiral media can cause asymmetric changes in the amplitudes of resonances of the CD spectra, without causing differential shifts in the position[17].

Reference measurements using achiral solutions were performed prior to experiments with chiral materials (supplementary). As expected there was no significant asymmetries between spectra from LH and RH structures, with $\Delta\Delta\lambda_{I,II,II}$ being ~ 0. It should be noted that the surface lattice (Bloch) mode III was more sensitive to the refractive index of the surrounding liquid the $\Delta\lambda_{L/R}$ being ~ 2 times greater than those for modes I and II.

**The chiral layers.**

An intrinsically chiral protein streptavidin has been used in this study, it is a tetramer with a predominately β-sheet structure[37]. Streptavidin was chosen because it can be utilised to produce both structurally isotropic and anisotropic chiral layers, the two cases referred to as specific and non-specific binding. If adsorbed directly from solution on to the nanostructures the protein adopts a broad range of orientations on the surface, characteristic of non-specific interactions, resulting in a layer with an isotropic structure. The small molecule biotin (sometimes referred to as vitamin $B_7$) binds very strongly to streptavidin, with a binding constant ($k_d$) ~ $10^{-14}$ $M^{38}$. This very strong interaction can be used to specifically bind streptavidin in a well-defined orientation. In particular, Au nanostructures were functionalised with self-assembled monolayers (SAMs) of a thiol with a biotin head group. Streptavidin specifically binds to these SAMs adopting a well-defined orientation[39]. It should be noted that biotin is also chiral, thus the SAMs will also be chiral. In addition to studying specifically and non-specifically bound streptavidin we have also made measurements from complexes formed by them and an antibody. Explicitly, a polyclonal mouse IgG which has been produced against streptavidin, subsequently referred to as anti-strept.

**CD data isotropic layers**

Spectra collected from the non-specific bound streptavidin layers are shown in **figure 5**. A red shift in the positions of the CD resonances occurs when unfunctionalized LH and RH structures are exposed to buffered solutions of streptavidin. This is consistent with an increase in the local refractive index around the nanostructures due to the adsorption of streptavidin. The spectra are not significantly changed after replacing the protein solution with buffer, indicating that the streptavidin is irreversibly adsorbed. The presence of the streptavidin induces no measurable asymmetry in the chiroptical properties, with $\Delta\Delta\lambda_{I, II, III}$ ~ 0. Binding anti-strept to streptavidin causes a further red shift in the CD resonances due to the increase in the thickness of the adsorbed layer. However, the (anti-strept)-streptavidin layer still does not cause a measurable asymmetry between CD spectra from LH and RH structures.

**CD data anisotropic layers**

The functionalisation of the gammadia with biotin SAMs causes a red shift in the spectra shown in **figure 6**, and there is a small but measurable asymmetry between CD spectra from LH and RH structures. Binding streptavidin to the biotin self-assembled monolayer (SAM) induces red shifts and results in a further measurable asymmetry between the gammadion enantiomorphs. Binding anti-strept to streptavidin induces a further red shift, and an increase in the level of asymmetries. A comparison between the values of $\Delta\Delta\lambda_{I, II, III}$ for the streptavidin and anti-strept depositions in the absence and presence of biotin, is shown in **figure 7**. They display a similar trend of mode II having the largest magnitude. However, the asymmetries for biotin have an opposite sign to those of streptavidin and (anti-strept)-streptavidin complex. The pattern and sign of asymmetries is similar to that observed in previous work of proteins adsorbed on to identical gammadia structures[28]. It should be noted that the smaller average shifts observed for binding of proteins to the SAM functionalised compared to the unfunctionalized structures can in part be attributed to the greater distance of the proteins from the surface. However, it is probable that there is a lower surface density of proteins in bound to the SAM, due to steric constraints inhibiting the biotin-streptavidin interaction. In addition to the asymmetric shifts in the three resonances there is also an asymmetry in the amplitude of the lattice resonance III for the immobilised streptavidin and anti-strept- streptavidin complex. This asymmetry in the amplitude of resonance III is also observed in previous gammadion studies for several non-specifically adsorbed proteins[28].

The experimental results clearly demonstrate the dependence on the asymmetry induced in the chiral layer and the level of structural anisotropy. It should also be noted that both the localised and lattice plasmon modes display asymmetries. The ability of structurally anisotropic layer to enhance the level of asymmetry between the (chir)optical response of LH and RH structures has been observed in previous studies involving a chiral metafilm based on six-armed shuriken structures. These structures exhibit chiroptical responses which are consistent with the helical oscillator model of chirality. The (chir)optical properties of these shuriken structures can be rationalised in terms of coupling between optically bright and dark modes[40]. In these cases, asymmetries in (chir)optical responses are reconciled in terms of coupling between modes being differentially modified by the presence of chiral dielectrics. The concept of chiral media causing asymmetric changes in coupling within the nanostructures, can be applied to the localised plasmonic modes of the gammadion structures. In this case the chiral dielectric perturbs the inductive coupling between orthogonal rods by causing a differential change in the chiral asymmetries / intensities of near fields occupying the gap region. Altering the level of coupling within the structure[31, 41] will result in a commensurate change in the far field chiroptical response. Consequently, the proposal is that the largest asymmetry in chiroptical response between LH and RH structures must be corelated to large differentials in the properties of near field regions between arms.

**Numerical simulations**

EM numerical simulations have been used to provide validation for the hypothesis proposed above. To accurately mimic protein layers we have defined dielectric slabs 20 nm thick, which cover each of the exterior surfaces of the gammadion, **figure 8.** The chiral properties of the dielectric slab are defined by $\xi$ a second rank complex tensor the sign of which is dependent on handedness, and is zero for achiral media. In the case where the electric dipole – magnetic dipole ($E_1M_1$) interaction is the dominant contributor to optical activity, then only the three diagonal elements $\xi_{xx, yy, zz}$ are non-zero. The interaction of EM fields with chiral media are given by the following constitutive equations:

$$\boldsymbol{D} = \varepsilon_o \varepsilon_r \boldsymbol{E} + i\xi \boldsymbol{B} \qquad (3)$$

$$\boldsymbol{H} = \boldsymbol{B}/\mu_0 \mu_r + i\xi \boldsymbol{E} \qquad (4)$$

Here, ($\varepsilon_r$) $\varepsilon_o$ is the (relative) permittivity of free space, and ($\mu_r$) $\mu_0$ is the (relative) permeability of free space. $\boldsymbol{E}$ is the complex electric field, and $\boldsymbol{H}$ is the magnetic field. Constitutive equations (3) and (4) were used in these simulations, and it was assumed that the chiral dielectric layers were continuous unstructured slabs.

To mimic the isotropic layers produced by the non-specific adsorption of proteins, the slabs were considered to have a homogeneous refractive index of 1.4, with a $|\xi| = 5 \times 10^{-4}$. When proteins are adsorbed on to a surface with a well-defined orientation, such as streptavidin *via* the biotin SAM, the properties of the layer are no longer isotropic. The refractive index in the direction of the surface normal will be different to those in the directions of the two orthogonal axes parallel to the surface, which are equal to each other due to azimuthal averaging. Thus, an oriented protein layers should be considered birefringent. It should be pointed out, that the anisotropy of chir(optical) properties in oriented proteins is due to the spatial distribution of the compound building blocks, rather than an intrinsic optical anisotropy of the building blocks themselves. We have simulated the birefringent layers with refractive index (n) components of $n_x = n_y = 1.3$ and $n_z = 1.6$, these values comparable to that previously measured for a protein system[42]. Simulated spectra for non-birefringent (isotropic) and birefringent (anisotropic) layers are displayed in **figures 9 and 10**.

The simulations for the isotropic layer display significant asymmetries in the amplitudes of modes I, II and III, but only the lattice mode III has a $\Delta\Delta\lambda \neq 0$. In contrast simulated spectra for the birefringent (anisotroptic) slabs are in better agreement with the experimental specifically bound data, with both $\Delta\Delta\lambda \neq 0$ values and asymmetries in the amplitudes for resonances II and III observable. A comparison between the $\Delta\Delta\lambda$ values derived from simulation and experiment are shown in **figure 7**.

**Optical chirality / Intensity maps.**

Maps showing the spatial distribution of intensities and $C$ of the EM fields, generated by incident left and right circularly polarised light (LCP and RCP) for resonances I, II and III are shown in **figures 11 and 12, respectively**. As expected for the achiral dielectric, symmetry related combinations of nanostructure handedness and light polarisation (*i.e.* LH/LCP = RH/RCP, LH/RCP = RH/LCP), give the same field intensity maps, and $C$ maps have opposite signs but are otherwise identical. However, the introduction of chiral dielectric breaks the symmetry relation for both field intensity and $C$ maps, with LH/LCP $\neq$ RH/RCP, LH/RCP $\neq$ RH/LCP. Table 1 contains averaged field and $C$ intensities for mode II, taken from regions between the arms of the gammadion. The level of breaking of the mirror relationships, is significantly greater for the birefringent slabs, with largest changes in both field intensity and C occurring in the region between the arms. We attribute the stronger influence of birefringent layers on near field properties to the ability to act as an additional sink / source of optical chirality. The birefringent layers have the largest differential effects on the near fields between the arms, with less pronounced differences in the fields between adjacent structures.

It is worth noting that the numerical simulations do underestimate the magnitude of the $\Delta\Delta\lambda$ for resonances I and II compared to experiment. This however is not surprising given that the near fields of the gap region would be very sensitive to the morphology / roughness /defect of the nanostructure which are not accounted for in the idealised model. For instance, hotspots associated with defects / roughness of the arms could strongly perturb the coupling.

**Conclusions**

The ability of birefringent layers to strongly perturb the chiral near fields is analogous to how they affect CPL. A beam of CPL propagating through a birefringent layer will suffer a level of depolarisation, becoming elliptically polarised, and reducing the $C$ of the beam. Alternatively, a linear polarised light beam, with polarisation that is not parallel to an optical axis will develop elliptical polarisation, hence gaining $C$. Consequently, (weakly absorbing) birefringent materials can act as a sink (or source) of optical chirality. The exchange of optical spin angular momentum from light to birefringent materials create opto-mechanical torques. In the current case, these torques would induce rotational motion of the immobilised protein molecules. Effectively, changes in $C$ of the near fields are due to the conversion of optical spin into molecular motion of the proteins.

This work provides insight in to the mechanism of enantiomeric sensing with gammadia, which could be generally applied to other structures. Numerical simulations suggest that chiral lattice modes are more sensitive to isotropic chiral media than localised modes. The greater sensitivity may be attributed to the fact that lattice modes derive from an ensemble of nanostructures, thus they sample a large amount of chiral material. Localised modes are less sensitive to chiral media, but the ability of birefringent chiral layers to perturb the properties of nearfields induces differential changes in resonances which are derived from coupling between structural elements. The sensitivity of these localised chiral resonances to the structure of the chiral media makes them ideally suited to the study of biomacromolecules, which generally are structurally anisotropic.

The complexity of biomolecules such as proteins introduce effects which will strongly influence the sensitivity of chiral plasmonic sensing. For instance, in solution proteins generally exist in polymeric form, with the number and spatial distribution of the individual monomer units controlling the overall structural anisotropy. Thus, the sensitivity of chiral plasmonic structures would be dependent on the aggregation state of the adsorbed proteins. If they are directly and non-specifically bound to a metal surface the level of aggregation would be controlled by several factors including: the type of protein; surface morphology / roughness of the nanostructure; and the ionic strength of the surrounding liquid. So, to some extent the aggregation state of the adsorbed protein will be a function of both how the sample was fabricated and its previous history, leading to variability and uncertainty in measurements. This variability can be drastically reduced, and the robustness of the methodology enhanced, by employing surface immobilisation techniques, such as the one used here, which were originally developed for biosensing applications. Additionally, the ability of point defects / surface roughness to influence the inductive coupling between structural elements will clearly affect the sensing capabilities of a chiral structure. Hence, the sensing efficiency would be expected to be dependent on anything that influences nanostructure morphology, such as metal deposition strategy used and whether the sample had been plasma cleaned.

In summary, by considering the ability of birefringent chiral layers to act as sinks / sources of optical chirality (optical spin angular momentum), the dependency of enantiomeric sensing on structural anisotropy can be understood. Our work highlights the potential strengths sensing with chiral near fields. In particular, the novel sensitivity to higher order biological structure of some chiral localised resonances which are derived from coupling between modes.

**Materials and Method**

**Gammadion Sample Fabrication**

The gammadia structures were fabricated using an electron beam lithography process. Quartz glass slides were cleaned under ultrasonic agitation in acetone, methanol and isopropyl alcohol (AMI) for 5 minutes each, dried under $N_2$ flow and exposed to $O_2$ plasma for 5minutes at 100W. A PMMA resist bilayer (*Allresist* 632.12 50K in Anisole and 649.04 200k in Ethyl Lactate) was then spun at 4000 rpm for 1 minute and baked at 180$^0$C for 5 minutes in between spins. A 10nm aluminium conducting layer was evaporated on the substrates using a *PLASSYS MEB 550s* evaporator. Patterns were designed on the CAD software *L-Edit* and written by a *Raith* EBPG 5200 electron beam tool operating at 100kV. The resist was developed in 3:1 MIBK:IPA solution at 23.2°C for 1 minute, rinsed in IPA (5s) and water before drying under $N_2$ flow. A 5nm nichrome adhesive layer was then evaporated below a 100nm gold layer. The process was completed with a lift-off procedure in acetone at 50°C overnight and then agitated to remove all remaining resist and excess metal.

**Sample Preparation**

Biotin-PEG-thiol ($C_{34}H_{65}O_{13}S$) was purchased from *Polypure* and dissolved in (Gibco) PBS buffer (10x, pH7.4) to a concentration of 60µM. Streptavidin protein was acquired from *Thermo Fisher* and diluted in PBS to a concentration of 2µM. Anti-streptavidin antibody produced in rabbit was obtained from *Sigma-Aldrich* and diluted in PBS to make up a 4µM solution.

Gammadion substrates were placed in a custom printed sample holder, with a *FastWell* Silicone seal and clear borosilicate glass slide above it. Solutions were injected through the seal. Samples were secured in a JASCO J-810 Spectropolarimeter to perform CD measurements. Biotin depositions were performed overnight to allow sufficient time for the SAM layer to form. The samples were then rinsed in PBS to remove unbound Biotin. Streptavidin solution in PBS was then introduced to the sample and left overnight, prior to rinsing in 0.1% NaOH/Tween which removed non-specifically bound streptavidin. Finally, antibody solution was injected and left to deposit for 2 hours, at which point a 0.05% NaOH/Tween solution was used to remove non-specifically bound antibody. Measurements were performed with biomolecule solutions and with PBS replacement. Samples were cleaned between experiments using AMI and a low power plasma clean.

**Numerical Simulations**

Simulations were performed using commercial finite element analysis software, COMSOL Multiphysics v5.4 (Wave optics module). Periodic boundary conditions were applied to emulate the array of structures. Perfectly matched layers were used above and below input and output ports. LCP- and RCP-light was applied at normal incidence through the quartz on to the gammadion. Additional 20nm 'protein' domains were extruded from the outer surface of each gammadion and split into discrete domains. The domains were identified by the axis of their surface normal (x, y or z). Protein domains were made chiral, with a $|\xi| = 5 \times 10^{-4}$. Protein domains were also assigned a refractive index of 1.4 in the isotropic case. For birefringent simulations, domains were given a birefringence of 1.3/1.6, with the largest values being assigned to the axial component of the respective surface normal.

**Acknowledgement**

The authors acknowledge financial support from the Engineering and Physical Sciences Research Council (EP/P00086X/1 and EP/M024423/1) Technical support from the James Watt Nanofabrication Centre (JWNC). DKM was awarded a studentship by the EPSRC. CG's work was supported by the EPSRC CDT in Intelligent Sensing and Measurement, Grant Number EP/L016753/1. MK acknowledges the Leverhulme Trust for the award of a Research Fellowship.

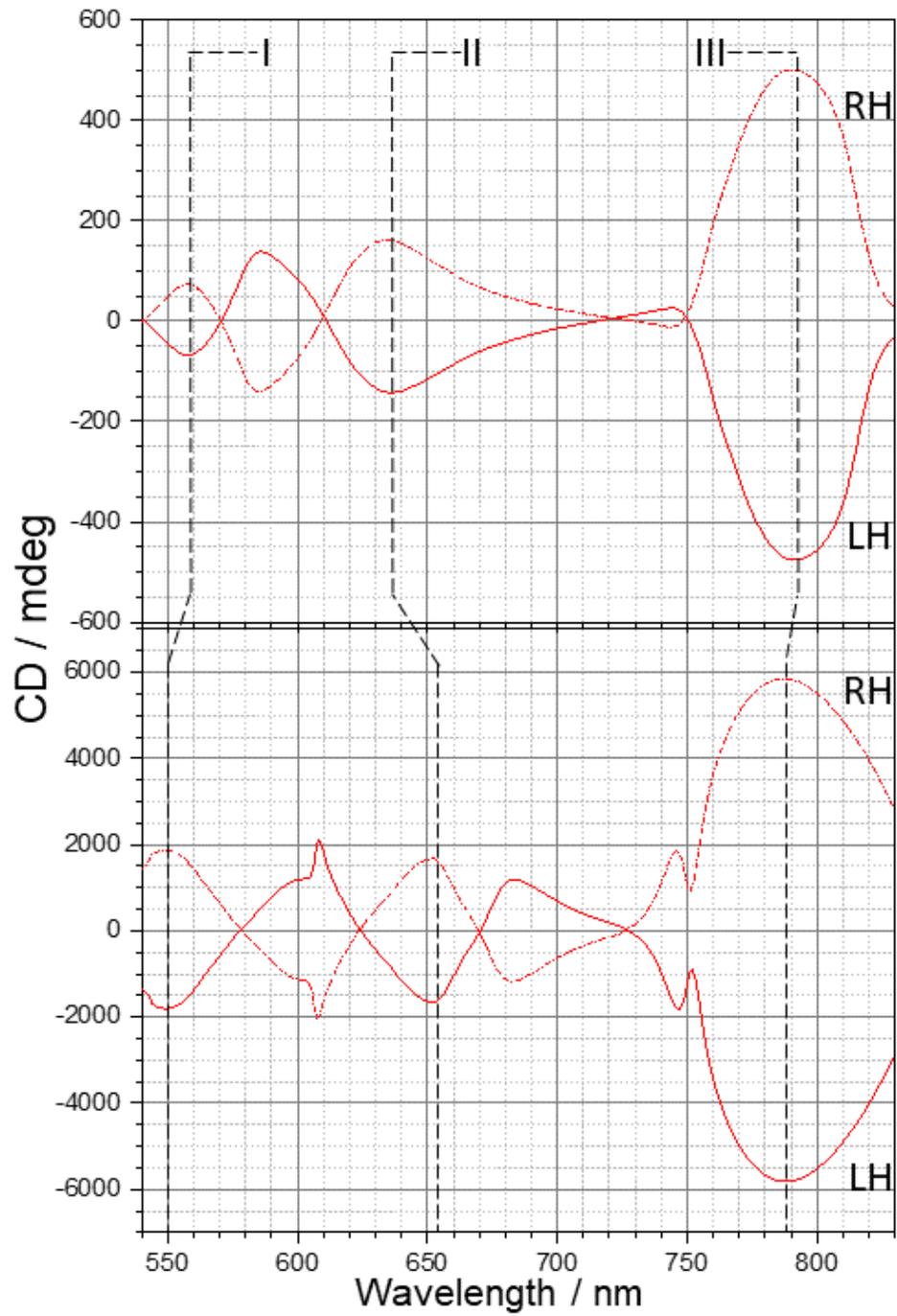

Figure 1: The experimental (top panel) and simulated (bottom panel) CD spectra of the two oppositely handed gammadion arrays (solid LH, dashed RH), modes I, II and III are labelled.

Figure 2: (a) An idealised RH gammadion geometry used for numerical simulation, with the dimensions labelled. (b) The 800nm periodic unit cell used in numerical simulations is shown. The gold gammadion lies upon quartz domains and has its outer surfaces exposed to water.(c) SEM images of the LH and RH gammadia used in experient.

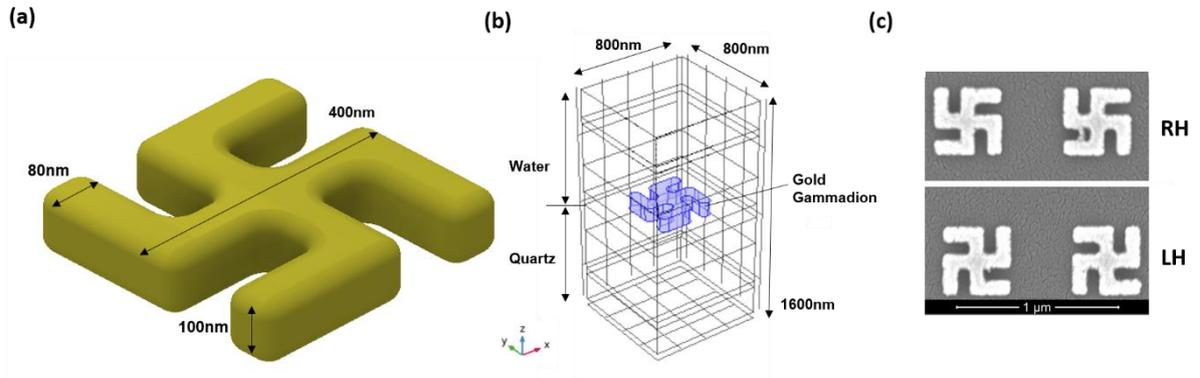

**Figure 3: (a) Electric field and (b) optical chirality plots for LH and RH (labelled in first column) gammadia structures in water, illuminated with RCP and LCP, for modes I, II and III are shown. Optical chirality plots are normalised to RCP.**

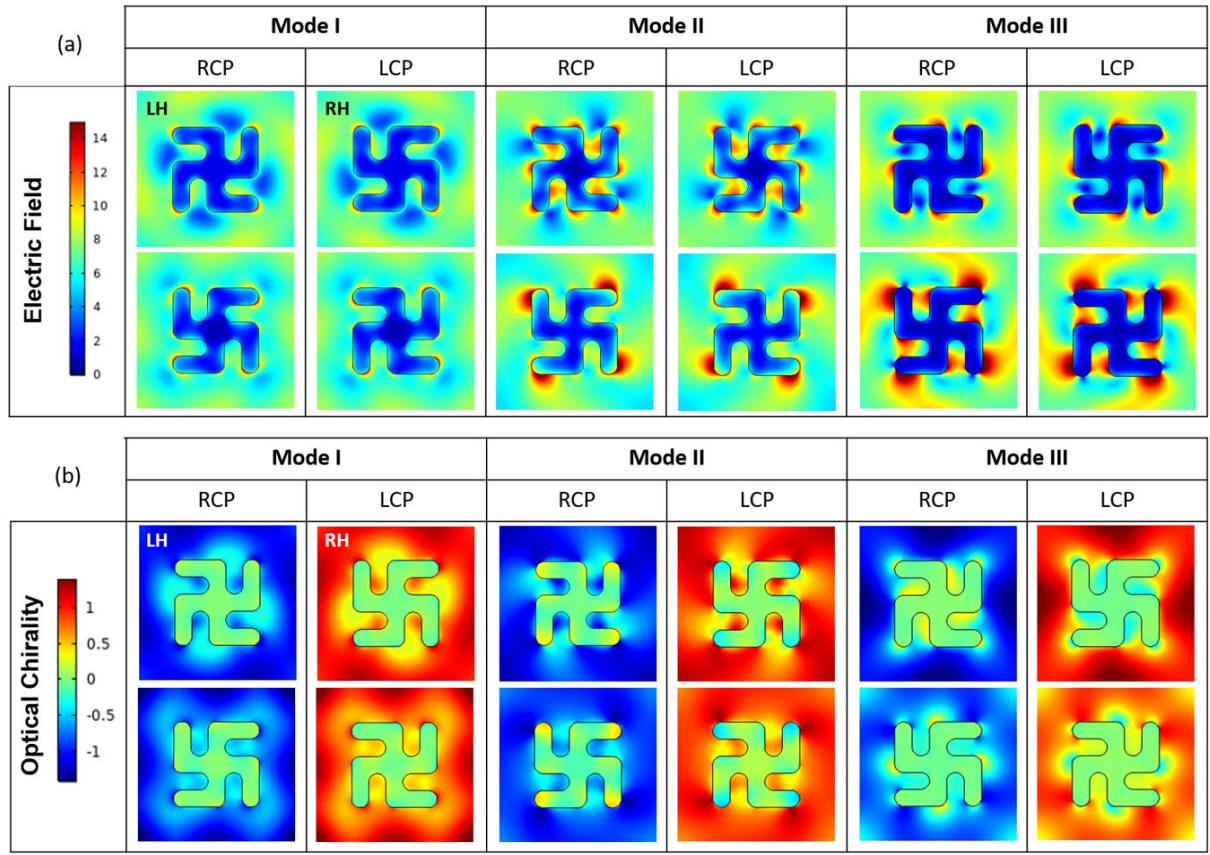

**Figure 4: Experimental CD spectra for LH (solid) and RH (dashed) gammadions in buffer solution when inclined at θ = 0° (red). 4° (black) and 10° (blue). The positions of mode I, II and III are labelled, the latter splits with increasing angle.**

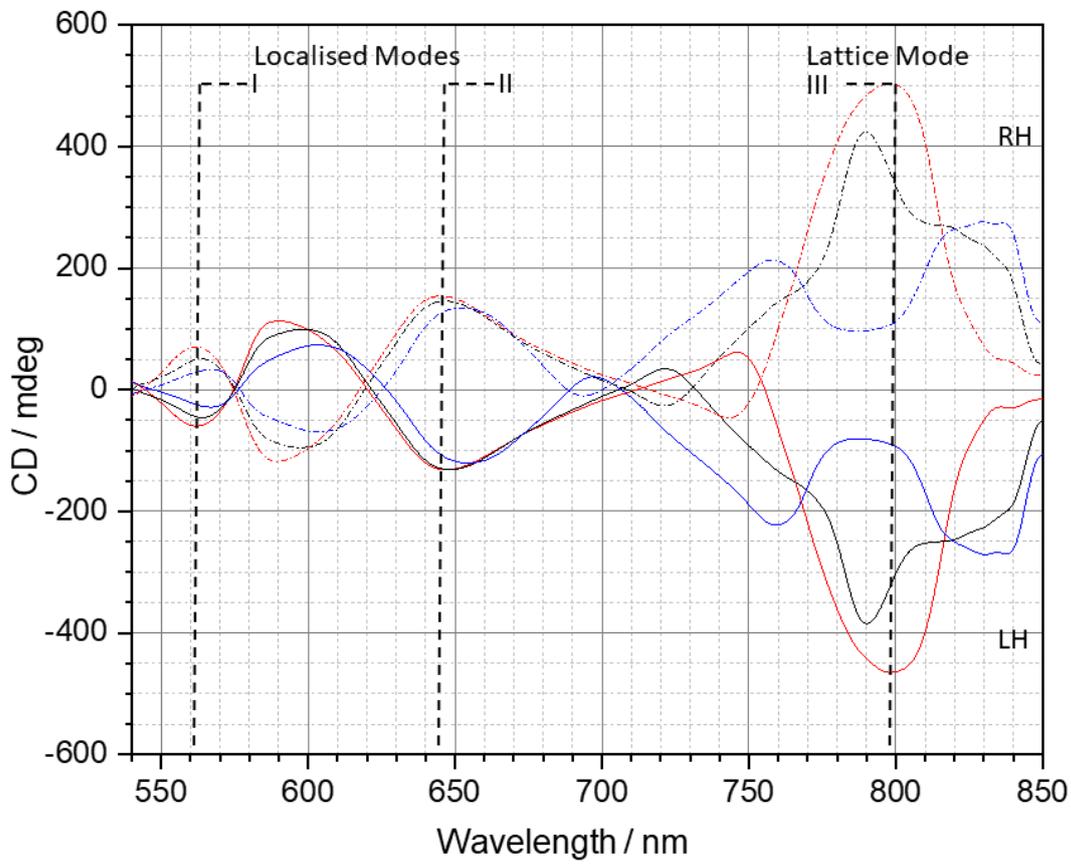
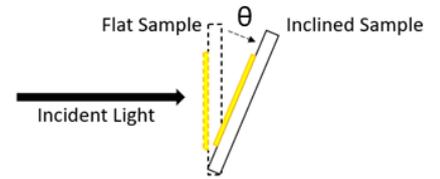

**Figure 5:** Experimental LH (solid) and RH (dash) CD spectra (top panel) for non-specific binding (isotropic) of streptavidin (black) with reference to PBS buffer (red). The subsequent addition of anti-strept (green) (bottom panel) with reference to streptavidin. Peak maxima for modes I, II and III are highlighted with lines to guide the eye (LH solid and RH dashed)

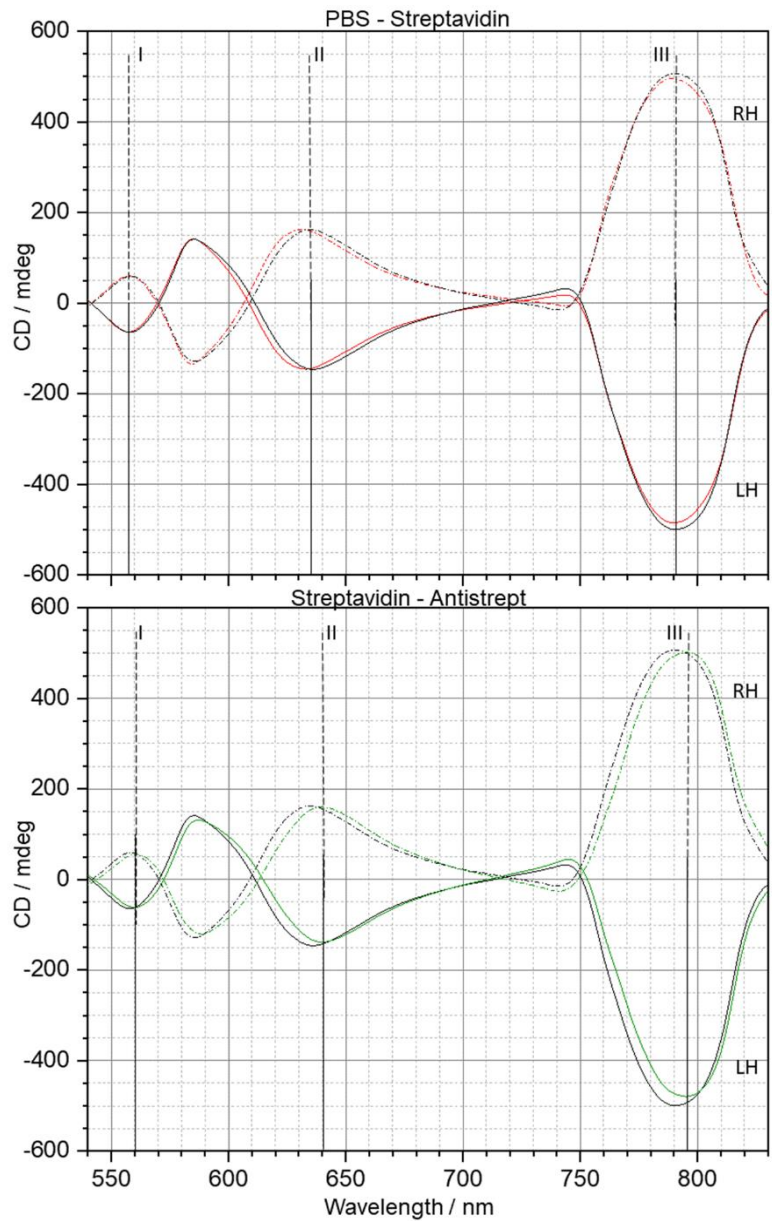

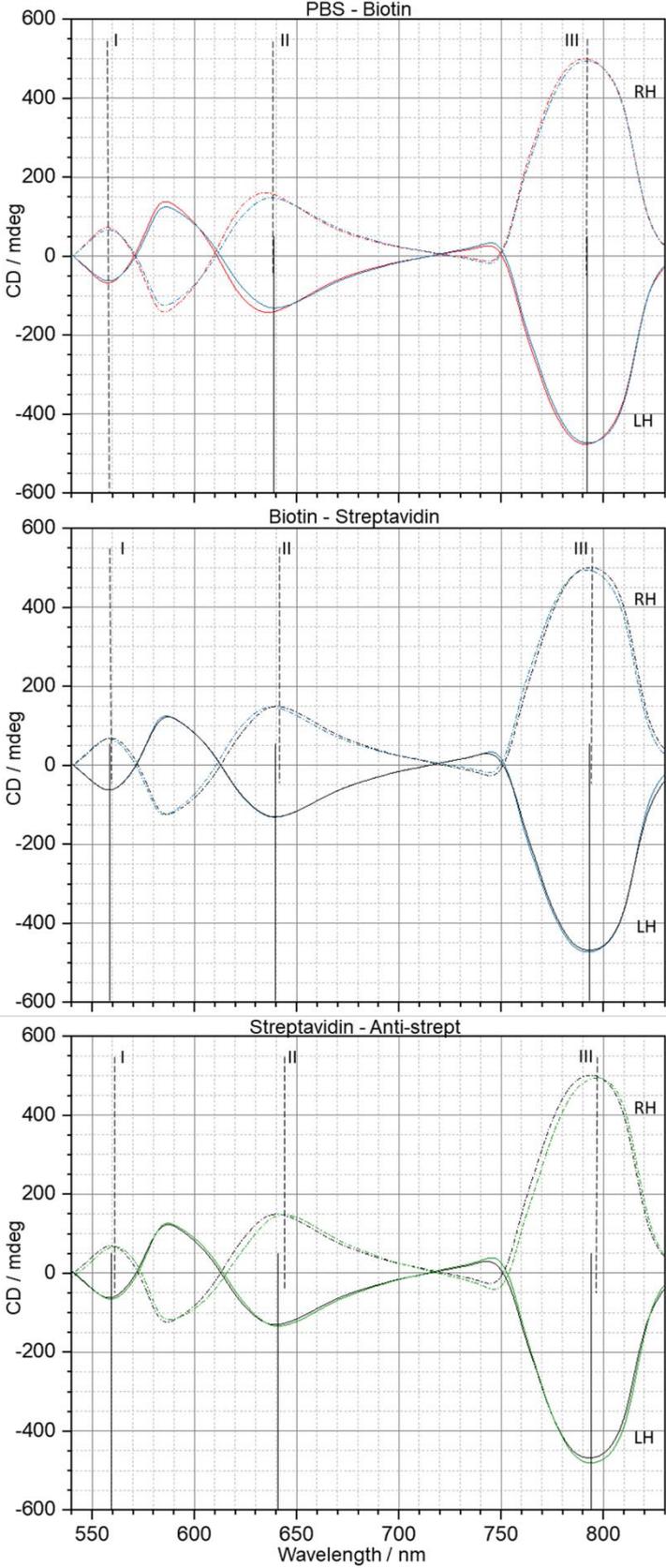

Figure 6: Experimental CD spectra for LH (solid) and RH (dashed) deposition of a biotin SAM (blue) with reference to PBS buffer (red), providing an anisotropic deposition profile of streptavidin (black) and anti-strept (green).

**Figure 7:** A comparison of the ΔΔλ asymmetry values for modes I (red), II (green) and III (blue) obtained from both experiment and simulation, for the isotropic and birefringent (anisotropic) layers.

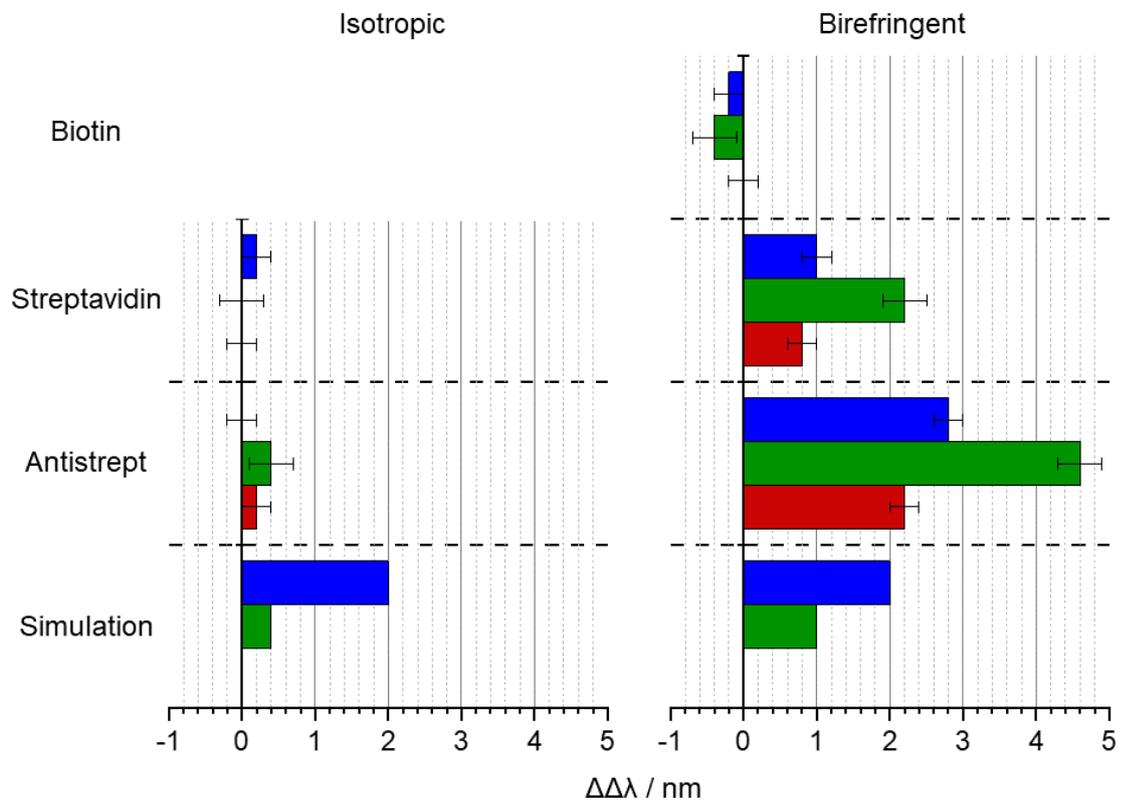

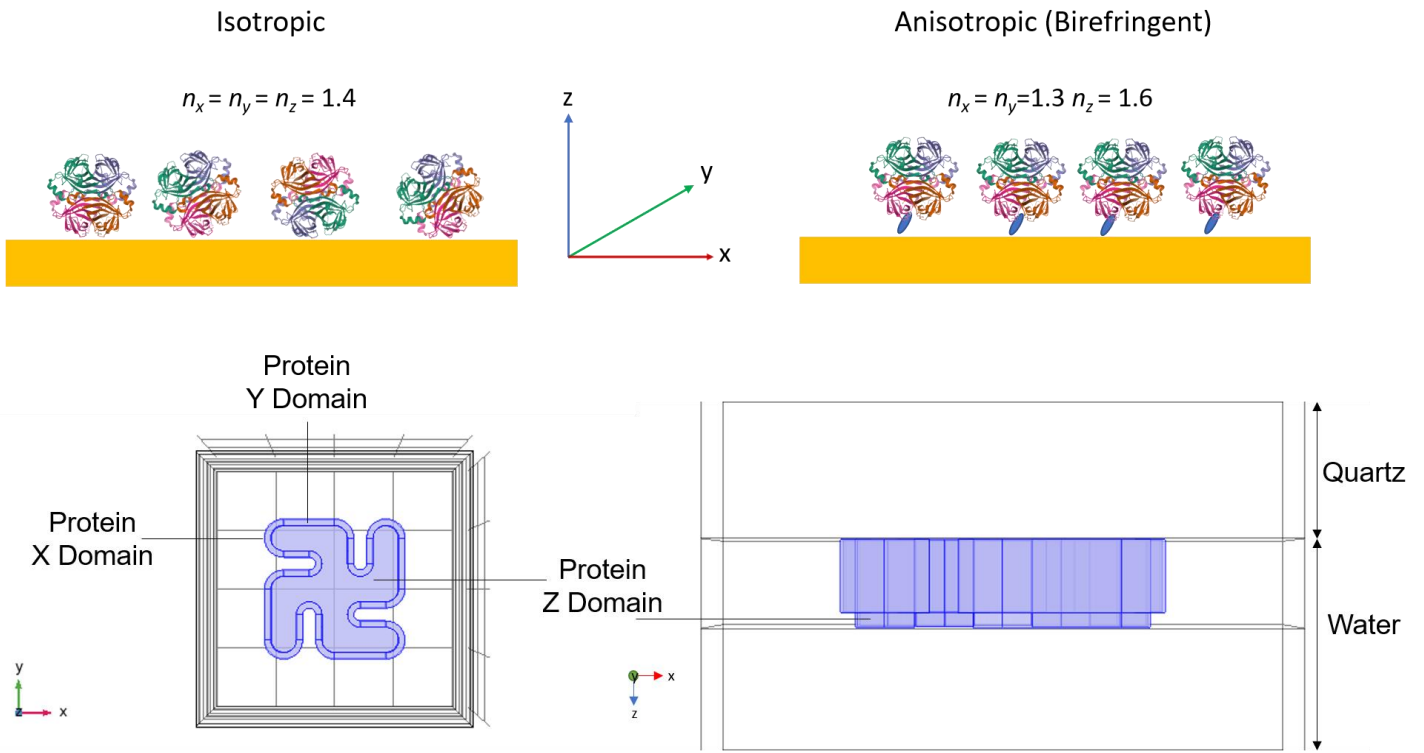

Figure 8: The upper panels shows a schematic diagram (not drawn to scale) illustrating the geometry of the streptavidin protein in non-specific (isotropic) and specifically (anisotropic) bound layers. The values for the refractive index (n) for the x,y and z component of the refractive index used in the simulations are given. In the lower panels the idealised model used to simulate the LH gammadion with a 20nm thick chiral dielectric layer is shown. The Dielectric domains are viewed from the top (left) and side (right) and are surrounded by water. For birefringent layers, the highest refractive index component is normal to the surface on which the domain is located.

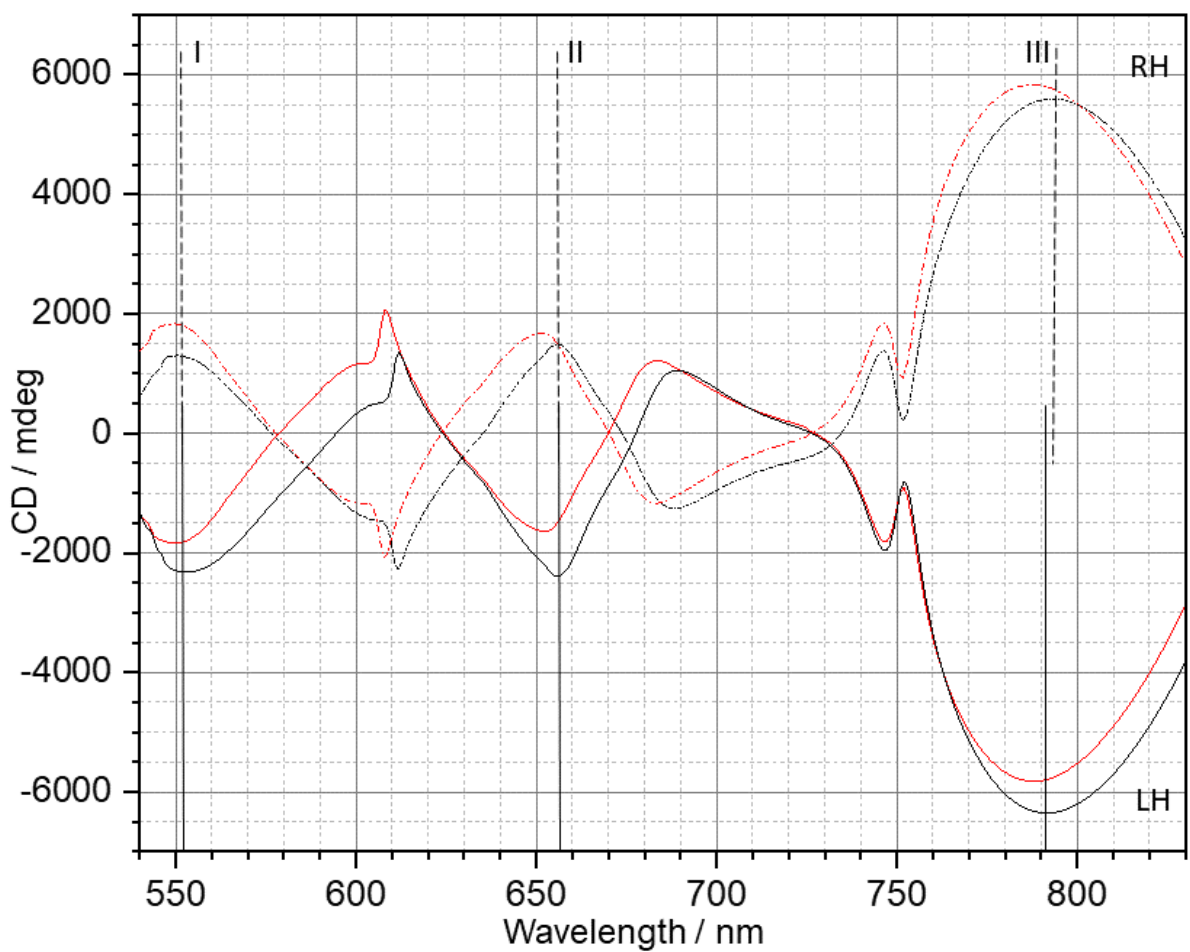

Figure 9: Simulated CD spectra for LH (solid) and RH (dashed) gammadia in water (red) and with an additional 20nm thick isotropic chiral dielectric layer (black).

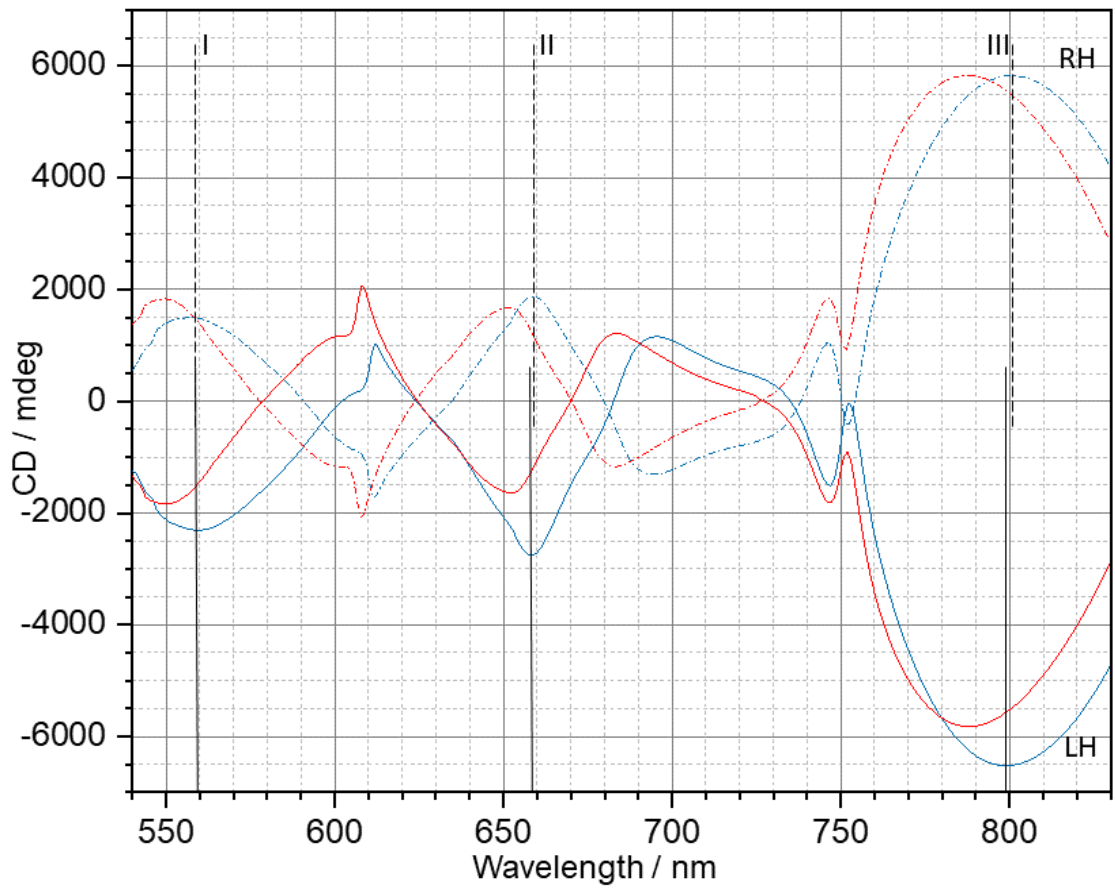

Figure 10: Simulated CD spectra for LH (solid) and RH (dashed) gammadia in water (red) and with an additional 20nm thick birefringent (anisotropic) chiral dielectric layer (blue).

**Figure 11:** Electric field plots of mode II for the achiral, isotropic and birefringent simulations. The intensities displayed in Table 1 are average values obtained from the black regions highlighted in achiral RCP / LH.

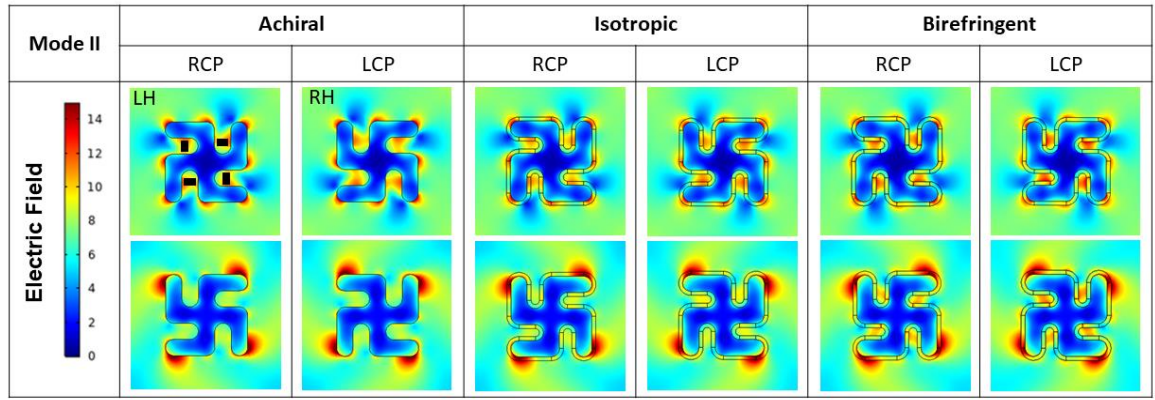

**Figure 12: Optical chirality plots of mode II for achiral, isotropic and birefringent simulations. The intensities displayed in Table 1 are average values obtained from the black regions highlighted in achiral RCP / LH.**

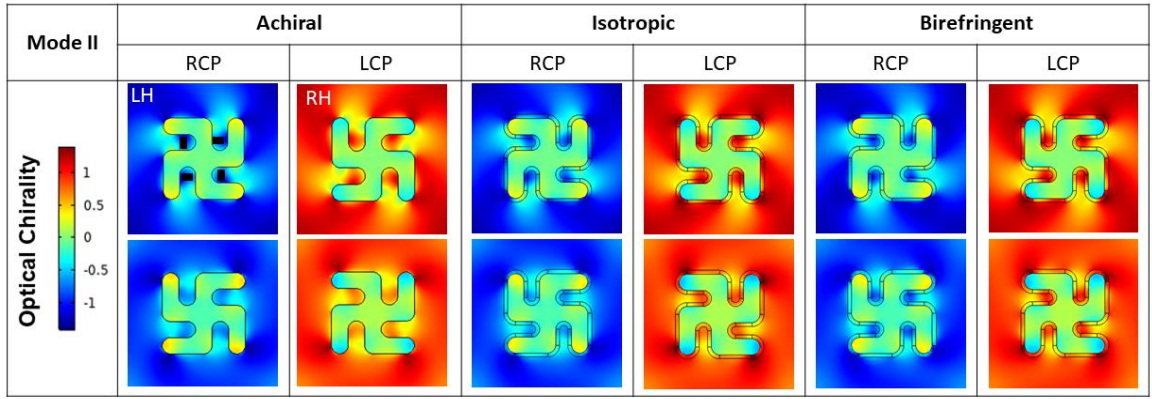

Table 1: A comparison of the electric field and optical chirality values, averaged from 4 equivalent areas between the four arms of the structures illustrated in figures 11 and 12. One set of symmetry equivalent handedness/polarisation pairs is shaded grey.

| Gammadion Handedness | Light Polarisation | Achiral | | Isotropic | | Birefringent | |
|---|---|---|---|---|---|---|---|
| | | Electric Field | Optical Chirality | Electric Field | Optical Chirality | Electric Field | Optical Chirality |
| LH | RCP | 10.09 | -0.72 | 10.06 | -0.64 | 11.10 | -0.73 |
| RH | LCP | 10.10 | 0.71 | 10.77 | 0.89 | 11.72 | 0.95 |
| LH | LCP | 8.33 | 0.54 | 9.21 | 0.69 | 10.76 | 0.82 |
| RH | RCP | 8.30 | -0.54 | 8.69 | -0.50 | 10.22 | -0.68 |